# Observation of Electronic Viscous Dissipation in Graphene Magneto-thermal Transport

Artem Talanov[1,2,]*, Jonah Waissman[1,3,]*, Aaron Hui[4], Brian Skinner[4], Kenji Watanabe[5], Takashi Taniguchi[6], Philip Kim[1,2]*

[1]Department of Physics, Harvard University, Cambridge, 02138, MA, USA
[2]John A. Paulson School of Engineering and Applied Sciences,
Harvard University, Cambridge, 02138, MA, USA
[3]Institute of Applied Physics, The Hebrew University of Jerusalem,
Jerusalem, Israel 9190401
[4]Department of Physics, The Ohio State University, Columbus, 43202, OH, USA
[5]Research Center for Electronic and Optical Materials,
National Institute for Materials Science, 1-1 Namiki, Tsukuba, 305-0044, Japan
[6]Research Center for Materials Nanoarchitectonics,
National Institute for Materials Science, 1-1 Namiki, Tsukuba, 305-0044, Japan

*Corresponding author. Email: philipkim@g.harvard.edu

---

*These authors contributed equally to this work

**Hydrodynamics describes the collective transport dynamics of particles as fluids with well-defined thermodynamic quantities. With enhanced electron-electron interactions at elevated temperatures, the collective behavior of electrons in graphene with minimal impurities can be depicted as a hydrodynamic flow of charges. In this new regime, the well-known rules of Ohmic transport based on a single electron picture no longer apply, necessitating the consideration of collective electron dynamics. In particular, the hydrodynamic analogues of Joule heating and thermal transport require consideration of the viscous motion of the electron fluid, which has a direct impact on internal energy dissipation and heat generation. In this work, we probe graphene hydrodynamics via thermal transport measurement and find two distinct, qualitative signatures of hydrodynamics: thermal conductivity suppression below the Wiedemann-Franz value, and viscous heating leading to a magnetic field dependence of the resultant temperature profile. We find these two effects are coincident in temperature and density despite arising from two distinct aspects of this new regime: microscopic momentum conservation due to electron-electron scattering, and geometry-dependent viscous dissipation. Our results mark the first observation of viscous electronic heating in an electron fluid, providing insight for thermal management in electronic hydrodynamic devices and offering a new methodology for identifying hydrodynamic states in other systems.**

# Introduction

Electrons moving in conventional conductors can be modeled as weakly-interacting quasiparticles, a concept described by Fermi liquid theory. Charge and energy transport in such systems often follow Ohm's law and the Wiedemann-Franz (WF) law, which links electrical and thermal transport (*1*). In the presence of strong electron-electron (e-e) interactions, this single-particle description can break down, often necessitating consideration of the collective motion of charge flow.

One strongly interacting electronic regime that has recently attracted significant attention is electron hydrodynamics, enabled by high e-e scattering rates and low sample disorder in specific materials (*2, 3*). In this regime, the e-e scattering rate exceeds all other scattering rates, causing electrons to behave as a low-viscosity fluid, with dynamics described by the Navier-Stokes equation using locally defined thermodynamic quantities, such as electronic temperature, pressure, and density. Monolayer graphene, an isolated single atomic layer of graphite with carbon atoms arranged in a hexagonal lattice, is particularly suitable for hydrodynamics due to its low disorder, negligible Umklapp scattering, and weak, quasi-elastic electron-phonon coupling (*4*).

Experimental evidence for electron hydrodynamics in graphene has accumulated, with experiments showing viscous backflow (*5, 6*), Poiseuille flow (*7, 8*), superballistic flow (*9, 10*), hydrodynamic constriction flow (*11*), Hall viscosity (*12*), vanishing of the Landauer-Sharvin resistance (*13*), and whirlpool imaging (*14*). These experiments focus on electrical transport and, so far, how hydrodynamic electrons dissipate heat has remained unaddressed. Furthermore, many of the observed electrical effects are subtle and can be difficult to discern from ballistic transport (*3, 6*) because e-e collisions conserve momentum and thus do not relax total electrical current. In addition, hydrodynamic phenomena like current vortices have been predicted and observed in non-hydrodynamic systems (*15, 16*), highlighting the need for new probes.

In contrast to electrical effects, thermal signatures have the potential to be stronger indicators of hydrodynamics due to decoupling of charge and energy currents. In certain highly-conductive bulk metals, thermal transport has recently been used to identify signatures of electronic hydro-dynamics (*17–20*). However, thermal transport in graphene is less experimentally studied due to the difficulty in measuring nanoscale thermal effects in low-dimensional systems. In graphene, previous thermal transport studies showed enhanced thermal transport in the charge-neutral regime of clean graphene (*21*), interpreted as the effect of an interacting Dirac fluid, and in the degenerate regime with strong disorder (*22*), interpreted as arising from quasi-elastic electron-flexural phonon coupling. To date, however, no study has conclusively demonstrated a clear thermal signature of the degenerate hydrodynamic electron fluid in graphene, such as viscous dissipation, where the kinetic energy converts to thermal energy via internal friction of the fluid.

In this work, we use Johnson noise thermometry and energy loss measurement to find two thermal signatures of a degenerate hydrodynamic liquid in graphene. First, we observe the predicted suppression of the thermal conductivity below the WF value (*4, 23–32*). Second, we observe negative thermal magneto-resistance arising from viscous heating due to particle current gradients, as opposed to Ohmic Joule heating due to impurity or phonon scattering (*33*). Our observations provide new insights applicable to studying magneto-hydrodynamic phenomena in strongly correlated electron liquids.

## Thermal Conductivity in Zero Magnetic Field

To measure thermal conductance, we use high-bandwidth Johnson noise thermometry to obtain the weighted spatial average of the temperature rise of the electron system in a 2-terminal (2T) graphene sample when it is heated by a quasi-DC current $I$ (*34–37*). The weighted-average temperature rise $\Delta T_{\text{avg}}$ is related to the thermal conductivity of the electrons through a generalized thermal conductance defined as

$$G_{\text{th,gen}} = \frac{I^2 R}{\Delta T_{\text{avg}}}, \tag{1}$$

where $I^2R$ is the total Joule heat generated in the sample with resistance $R$ (*38*) (see Supplementary Materials (SM)). This quantity is distinct from but related to the linear-response thermal conductance obtained when applying a temperature gradient across a sample, since in linear response there is no dissipation in the measured sample, only energy current, and hence no dependence on $R$ (*39*). By comparing $G_{\text{th,gen}}$ with the electrical conductance $G_{\text{el}} = 1/R$, we obtain the sample Lorenz ratio $\mathcal{L}/\mathcal{L}_0$, which relates thermal conductivity $\kappa$ to electrical conductivity $\sigma$ for Ohmic transport in an arbitrarily shaped 2T sample via a universal factor (*38*):

$$\frac{\mathcal{L}}{\mathcal{L}_0} = \frac{1}{\mathcal{L}_0 T} \times \frac{\kappa}{\sigma} = \frac{1}{\mathcal{L}_0 T_0} \times \frac{G_{\text{th,gen}}}{12 G_{\text{el}}}, \tag{2}$$

where $\mathcal{L}_0$ is the Sommerfeld value of the Lorenz number, given by $\mathcal{L}_0 = \frac{\pi^2}{3}(\frac{k_\text{B}}{e})^2$, with $k_\text{B}$ the Boltzmann constant, $e$ the electron charge, and $T$ the temperature. The WF law corresponds to $\mathcal{L}/\mathcal{L}_0 = 1$.

The insets in Fig. 1(a,b) show the 2T Corbino geometry we use for this experiment (see SM for device fabrication details). We first characterize the electrical and thermal properties of the device under zero magnetic field. Fig. 1(a) shows the measured two-terminal electrical resistance $R_{\text{el}}$ vs electron density $n$ for three representative temperatures: 24, 90, and 200 K. In this device, for $T \lesssim 90$ K the Dirac peak width becomes disorder-limited with a residual density of $\sim 5.4 \times 10^{10}$ cm$^{-2}$ (see SM). Figure 1(b) shows the measured thermal resistance, $R_{\text{th}} = \mathcal{L}_0 T/G_{\text{th,gen}}$, normalized to units of $\Omega$ to facilitate a comparison with $R_{\text{el}}$. At 24 K, the thermal resistance qualitatively matches the electrical resistance. As temperature is increased, the thermal resistance develops a dip near $n = 0$ and one peak on each side of charge neutrality, features which are absent in the electrical resistance.

The Lorenz ratio is shown in Figure 1(c). At 24 K, the Lorenz ratio near charge neutrality is close to 1 and gate-independent (see Fig. 3a for the full WF regime window). At higher density, the measured Lorenz ratio increases above 1, owing to the transition to ballistic transport where Eq. 1 no longer holds (see SM).

At 90 K there develops a small peak at charge neutrality in the measured Lorenz ratio, decreasing to a slightly smaller value by 200 K. The magnitude of the peak is consistent with both a weak hydrodynamic enhancement (*24–26, 28, 30, 31, 40*) and a bipolar diffusion enhancement (*41, 42*), but our experiment cannot measure their relative contributions. Typically, bipolar diffusion effects grow with temperature, but this peak weakens from 90 K to 200 K, consistent with a hydrodynamic enhancement at 90 K getting suppressed by additional phonon scattering at higher temperatures, as observed in Ref. (*21*). Here, the value of the enhancement is lower and occurs at a different temperature, which could be explained by $\sim 10\times$ larger disorder in this device compared to the cleanest device in Ref. (*21*).

Concurrently with the $n = 0$ peak at 90 K, a relative suppression in the Lorenz ratio develops at a density of $n \sim \pm 1.1 \times 10^{11} \text{cm}^{-2}$. At 200 K the Lorenz ratio goes below 1, with the lowest measured value being $\sim$0.45.

The observed Lorenz suppression may be caused by inelastic phonon scattering (*43*), but only for $T \lesssim T_{BG} \sim 18$ K at this density, where the Bloch-Grüneisen temperature $T_{BG} = 2\hbar v_s k_F / k_B$, with $v_s$ the speed of sound, and $k_F$ the Fermi wavevector. Lorenz suppression is also possible due to thermal smearing of the Fermi surface, for $T \gtrsim T_F \sim 450$ K at this density, where $T_F$ is the Fermi temperature (*43*). However, the observed suppression occurs in the degenerate electronic regime and in the quasi-elastic electron-phonon regime throughout the measured temperature range (see Fig. 3(a)), ruling out known single-particle effects as causing the suppression. The Corbino geometry also rules out thermoelectric contributions (see SM). We emphasize that energy loss to phonons in a self-heating measurement increases $G_{\text{th,gen}}$ (*36*); thus, coupling to phonons cannot explain the observed increase in $R_{\text{th}}$ with $T$ for $|n| > 0$ and resulting Lorenz suppression.

In contrast, hydrodynamic relaxation of thermal current via e-e collisions (*23, 27*) is con-sistent with our observations. This suppression occurs in the hydrodynamic limit $\gamma_{ee} \gtrsim \gamma_{MR}$, where $\gamma_{ee}$ is the e-e scattering rate, and $\gamma_{MR}$ is the momentum-relaxing scattering rate. Furthermore, the Lorenz suppression occurs in a temperature and density range consistent with those measured in other electrical-only hydrodynamic experiments (*5, 9, 13*). However, the apparent Lorenz enhancement at high densities due to the transition to ballistic transport (see SM) complicates quantitative analysis of the scattering rates.

## Electrical and Thermal Magneto-Resistance

To further investigate this Lorenz suppression at finite density, we study the evolution of electrical and thermal transport in a perpendicular magnetic field to realize an independent measure of hydrodynamic parameters. For diffusive transport, the Drude model predicts the electrical conductivity decreases in a magnetic field $B$ as

$$\sigma_{xx}(B) = \frac{\sigma(0)}{1 + (\mu B)^2}, \qquad (3)$$

where $\sigma(0)$ is the conductivity at zero magnetic field, and $\mu$ is the density-dependent electrical mobility. The 2T resistance of a Corbino device is related to the conductivity via

$$R_{\text{el}}(B) = R_c + \frac{1}{\sigma_{xx}(B)} \frac{\ln(r_o/r_i)}{2\pi}, \qquad (4)$$

where $R_c$ is the contact resistance that we approximate to be independent of magnetic field for low fields (*44*), and $r_o$ and $r_i$ are the outer and inner contact radii, respectively. Hence, a simple diffusive model predicts a parabolic dependence of resistance on magnetic field.

Figure 2(a,c,e) shows the electrical magnetoresistance (EMR) $\Delta R = R_{\text{el}}(B) - R_{\text{el}}(0)$ for three different electron densities, for $T = 15 - 200\,\text{K}$. The measured EMR is positive and parabolic upwards to the highest measured magnetic field of 200 mT, as expected from the Drude picture.

For direct comparison with the EMR, we consider the thermal magnetoresistance (TMR) $\Delta R_{\text{th}} = R_{\text{th}}(B) - R_{\text{th}}(0)$, shown in Fig. 2(b,d,f) for the same three densities and the same temperature range of the corresponding EMR. Remarkably, in contrast to the EMR, the TMR changes sign from positive to negative as a function of temperature, density, and magnetic field. At $T = 15$ and $24$ K, the TMR closely follows the EMR, while at higher temperatures from 90-200 K the TMR becomes negative for small $B$, returning to positive values at higher $B$, with the strongest negativity occurring at densities $|n| > 0$.

To quantitatively characterize the low-$B$ EMR and TMR, we fit parabolas of the form $\Delta R_{el,th} = A_{el,th} B^2$ for data up to $B \sim 30$ mT. Example fits are shown in Fig. 2(g and f). The resulting EMR and TMR fit coefficients $A_{\text{el}}$ (black) and $A_{\text{th}}$ (dark green), respectively, are plotted in Fig. 2(i) for 200 K, overlaid with the measured Lorenz ratio (red, right-hand axis) at $B = 0$, reproduced from Fig. 1(c). At 200 K, $A_{\text{th}}$ is negative for all measured $n$, attaining the most negative value at $n = 1.3 \times 10^{11}$ cm$^{-2}$. This closely corresponds to the density where the Lorenz ratio at $B = 0$ is most suppressed, indicating a correlation between these two effects.

We now compare the observations over the entire density and temperature range of the measurement. Figure 3(a) shows a 2D plot of the Lorenz ratio as a function of $n$ and $T$, while Fig. 3(b) shows a 2D plot of the ratio $A_{\text{th}}/A_{\text{el}}$, for which deviations from 1 may indicate non-trivial transport regimes. We emphasize that $A_{\text{th}}/A_{\text{el}}$ depends only on the relative changes of $R_{\text{el}}(B)$ and $R_{\text{th}}(B)$ and is independent of the $B = 0$ Lorenz ratio. Nevertheless, we find that there is a strong correlation between the region of the Lorenz suppression ($\mathcal{L}/\mathcal{L}_0 < 1$), marked by a dark red colored region in Fig. 3(a), and the region of negative TMR ($A_{\text{th}}/A_{\text{el}} < 0$), marked by dark green color in Fig. 3(b). The conjunction of these effects in two different experimental quantities suggests related origins.

We can exclude single-particle explanations by comparing $T$ with characteristic energy scales. In Fig. 3, we plot the Fermi temperature, the Bloch-Grüneisen temperature (*45*), and the temperature scale of graphene Landau level spacing given by $k_\text{B}T = \hbar\omega_c$ for $B = 30$ mT, where $\omega_c$ is the cyclotron frequency. The Lorenz ratio suppression and the negative TMR both occur in the degenerate regime for $T \ll T_F$, and at high enough temperatures, $k_\text{B}T \gg \hbar\omega_c, T_{BG}$, that avoid both Landau quantization and inelastic phonon scattering. Hence, these effects are ruled out as the origins of both the $B = 0$ Lorenz suppression and the negative TMR.

## Viscous Heating in Hydrodynamics

The overlap of the negative TMR region with the hydrodynamic Lorenz suppression indicates that hydrodynamic effects under magnetic fields need to be considered. We will show that viscous heating explains the observed negative TMR. As a starting point, we first examine the non-viscous Ohmic regime. Here, heating of the electrons arises from Joule dissipation via impurity or phonon scattering, with a power density $p$ given by

$$p = \vec{J} \cdot \vec{E}, \tag{5}$$

where $\vec{J}$ is the electrical current density and $\vec{E}$ is the electric field. In our Corbino devices where the distance $r$ is measured from the center of the device, both components scale as $\propto 1/r$, giving a $\propto 1/r^2$ power dissipation profile, independent of magnetic field, with the resulting temperature profile also independent of magnetic field. Thus, the generalized thermal resistance is

$$R_\text{th} = \frac{\Delta T_\text{avg}}{P} = \frac{1}{12\kappa_{xx}}\frac{\ln(r_o/r_i)}{2\pi}, \tag{6}$$

and depends on $B$ only through $\kappa_{xx}$. The Ohmic heating and resulting temperature profiles are shown in Fig. 4(a,b,e,f,g,j).

In contrast, in the viscous regime heat dissipation arises from viscous shearing, the deformation of fluid elements caused by velocity gradients. In a Corbino geometry at $B = 0$, viscous electron flow has zero electric field inside the channel, as there is no momentum relaxation, and fluid elements experience zero net viscous force (*13, 46*). This contrasts with Poiseuille flow in a rectangle, where momentum relaxation comes from no-slip boundary conditions along the walls and fluid elements do experience a net viscous force with a non-vanishing electric field (*33*). Although the viscous velocity field at $B = 0$ is the same as in the Ohmic case ($\propto 1/r$), the heating instead arises from viscous dissipation and scales as $\propto 1/r^4$ due to velocity gradients in the viscous stress tensor (*33*). The heating profile and temperature distributions are hence concentrated closer to the inner electrode, as seen in Fig. 4(c,h).

Upon application of an external magnetic field, the electron fluid acquires a rotational velocity due to the Lorentz force in both the Ohmic and viscous regimes, as shown in Fig. 4 (b,d). In the Ohmic case, the power dissipation maintains a $\propto 1/r^2$ profile due to the radial-only electric field and $\propto 1/r$ radial velocity component.

In contrast, in the viscous case, the rotational component of velocity, which depends on $B$, now contributes to the power dissipation. The no-slip boundary condition at the electrodes generates strong velocity gradients near the electrodes, thus concentrating additional power dissipation near the electrodes (*46–48*). For a strong enough magnetic field, the viscous stress tensor is maximal near the electrodes with a minimum at an intermediate point along the channel. The shear viscous dissipation is proportional to the square of the viscous stress tensor and thus also acquires a minimum in the channel, as seen in the color map of Fig. 4(d), and in the normalized heating profile in Fig. 4(e). Compared to the Ohmic case, the resulting temperature distribution has a lower peak, visible in the color map Fig. 4(i), and in the normalized temperature profile in Fig. 4(j).

The reshaping of the temperature profile due to viscosity can act to decrease the measured $R_{\text{th}}$ for a fixed $\kappa_{xx}$. Our theoretical analysis shows that viscous dissipation and hydrodynamic noise generation modifies the generalized thermal resistance depending on $\nu$, the kinematic vis-cosity, and $\gamma_{MR}$, the momentum-relaxing scattering rate (*48*). The resulting thermal resistance possesses two limits: fully Ohmic, in which $\nu \to 0$, and fully viscous, in which $\lambda \to \infty$, where the Corbino device viscous parameter $\lambda = (l_G/r_i)$, with the radius of the inner contact $r_i$ and the Gurzhi length $l_G = \sqrt{\nu/\gamma_{MR}}$. For sufficiently large $\lambda$, $R_{\text{th}}(B)$ can decrease due to viscous reshaping of the heating profiles in the Corbino geometry, resulting in a temperature profile with reduced total thermal noise (Fig. 4(j)).

We now show that our theory and the combined electrical and thermal data allow for a quantitative measure of electronic viscosity and a new parameter, the thermal mobility $\alpha_{\text{th}}$. In analogy to Eq. 3, we assume $\kappa_{xx}(B) \propto 1/(1 + (\alpha_{\text{th}}B)^2)$ and use the combined electrical and thermal data sets to constrain $\alpha_{\text{th}}$, $\nu$, and $\gamma_{MR}$. In this approach, the small-field decrease in $R_{\text{th}}(B)$ is captured by our viscous heating theory, discussed above, while the large-field increase arises from the parabolic $\kappa_{xx}(B)$. Figure 5(a) shows EMR and TMR data sets at 200 K and varying $n$, with an exemplary set of fits, demonstrating how the concurrent positive EMR and negative TMR is captured by our theory. The resulting values of $\alpha_{\text{th}}$, the electrical mobility $\mu$ (related to $\gamma_{MR}$), and $\nu$ are shown in Fig. 5(b)-(d). The thermal mobility $\alpha_{\text{th}}$ exhibits a dip aligned with the $B = 0$ Lorenz ratio suppression, showing that zero- and finite-field data are mutually consistent with hydrodynamic suppression of thermal transport. The viscosity $\nu$, approximately in the range of $0.2 - 1.4$ m$^2$/s, is larger than that obtained in electrical-only experiments (*5, 8, 9*). It also shows a decreasing $1/n$ density-dependence, comparable to a previously-observed decreasing behavior (*5*). We emphasize that our approach does not depend on boundary scattering effects, geometric details (only the contact radii), or Ohmic contribution subtractions, and hence provides a new route to measure the electronic viscosity.

# Conclusion

In conclusion, we observe hydrodynamic suppression of the Lorenz ratio coinciding with negative thermal magnetoresistance in monolayer graphene Corbino devices. These effects arise from two aspects of the electronic hydrodynamic regime: 1) momentum-conserving e-e scattering that decays thermal currents, and 2) spatial redistribution of viscous heating under a magnetic field. Negative thermal magnetoresistance, together with positive electrical magnetoresistance, is a new, qualitative signature for the viscous magneto-hydrodynamic regime. Together with Lorenz suppression at zero magnetic field, these two features can distinguish the viscous and non-viscous hydrodynamic regimes. Importantly, these measurements utilize a simple, two-terminal, global transport setup, greatly simplifying experimental conditions for identifying electronic hydrodynamics, including in realistic device scenarios. They also offer new quantitative estimates of the fluid viscosity that are robust to details that have affected previous electrical measurements. Our work provides quantitative evidence for viscous heating and hydrodynamic thermal transport in graphene electronic hydrodynamic devices, providing insight into heat dissipation in this electrical and thermal transport regime. Furthermore, these methods can be straightforwardly applied to a diverse range of other systems in the search for hydrodynamic behaviors, including 2D systems like twisted graphene and bulk 3D systems.

# References

1. N. W. Ashcroft, N. D. Mermin, *Solid State Physics* (Holt, Rinehart and Winston, 1976).

2. M. Polini, A. K. Geim, *Physics Today* **73**, 28 (2020).

3. L. Fritz, T. Scaffidi, *Annual Review of Condensed Matter Physics* **15**, 17 (2024).

4. A. Lucas, K. C. Fong, *Journal of Physics: Condensed Matter* **30**, 053001 (2018).

5. D. A. Bandurin, *et al.*, *Science* **351**, 1055 (2016).

6. D. A. Bandurin, *et al.*, *Nature Communications* **9** (2018).

7. J. A. Sulpizio, *et al.*, *Nature* **576**, 75 (2019).

8. M. J. H. Ku, *et al.*, *Nature* **583**, 537 (2020).

9. R. K. Kumar, *et al.*, *Nature Physics* **13**, 1182 (2017).

10. Z. J. Krebs, *et al.*, *Science* **379**, 671 (2023).

11. A. Jenkins, *et al.*, *Phys. Rev. Lett.* **129**, 087701 (2022).

12. A. I. Berdyugin, *et al.*, *Science* **364**, 162 (2019).

13. C. Kumar, *et al.*, *Nature* **609**, 276 (2022).

14. M. Palm, *et al.*, *Science* **384**, 465 (2024).

15. A. Aharon-Steinberg, *et al.*, *Nature* **607** (2022).

16. J. Estrada-Alvarez, F. Domınguez-Adame, E. Dıaz, Alternative routes to electron hydrodynamics (2023).

17. J. Gooth, *et al.*, *Nature Communications* **9** (2018).

18. A. Jaoui, *et al.*, *Nature Communications* **3** (2018).

19. A. Jaoui, B. Fauqué, K. Behnia, *Nature Communications* **12** (2021).

20. W. Xie, *et al.*, Purity-dependent lorenz number, electron hydrodynamics and electron-phonon coupling in wte2 (2023).

21. J. Crossno, *et al.*, *Science* **351**, 1058 (2016).

22. M. M. Sadeghi, *et al.*, *Nature* **617**, 282 (2023).

23. A. Principi, G. Vignale, *Phys. Rev. Lett.* **115**, 056603 (2015).

24. A. Lucas, J. Crossno, K. C. Fong, P. Kim, S. Sachdev, *Physical Review B* **93** (2016).

25. H.-Y. Xie, M. S. Foster, *Physical Review B* **93** (2016).

26. Y. Seo, G. Song, C. Park, S.-J. Sin, *Journal of High Energy Physics* **2017** (2017).

27. A. Lucas, S. D. Sarma, *Physical Review B* **97** (2018).

28. M. Zarenia, A. Principi, G. Vignale, *2D Materials* **6**, 035024 (2019).

29. M. Zarenia, T. B. Smith, A. Principi, G. Vignale, *Physical Review B* **99** (2019).

30. S. Li, A. Levchenko, A. V. Andreev, *Physical Review B* **102** (2020).

31. S. Li, A. Levchenko, A. V. Andreev, *Physical Review B* **105** (2022).

32. S. Ahn, S. Das Sarma, *Phys. Rev. B* **106**, L081303 (2022).

33. A. Hui, B. Skinner, *Phys. Rev. Lett.* **130**, 256301 (2023).

34. K. C. Fong, K. C. Schwab, *Physical Review X* **2** (2012).

35. K. C. Fong, *et al.*, *Physical Review X* **3** (2013).

36. J. Crossno, X. Liu, T. A. Ohki, P. Kim, K. C. Fong, *Applied Physics Letters* **106**, 023121 (2015).

37. A. V. Talanov, J. Waissman, T. Taniguchi, K. Watanabe, P. Kim, *Review of Scientific Instruments* **92**, 014904 (2021).

38. C. Pozderac, B. Skinner, *Phys. Rev. B* **104**, L161403 (2021).

39. J. Waissman, *et al.*, *Nature Nanotechnology* **17** (2022).

40. A. Lucas, S. D. Sarma, *Physical Review B* **97** (2018).

41. H. Yoshino, K. Murata, *Journal of the Physical Society of Japan* **84**, 024601 (2015).

42. Y.-T. Tu, S. D. Sarma, *Physical Review B* **107** (2023).

43. A. Lavasani, D. Bulmash, S. D. Sarma, *Physical Review B* **99** (2019).

44. M. Kamada, *et al.*, *Physical Review B* **104** (2021).

45. D. K. Efetov, P. Kim, *Physical Review Letters* **105** (2010).

46. M. Shavit, A. Shytov, G. Falkovich, *Physical Review Letters* **123** (2019).

47. A. Levchenko, J. Schmalian, *Annals of Physics* **419**, 168218 (2020).

48. A. Hui, *arxiv* (2023).

# Acknowledgments

We thank Sang-Jin Sin, Andrew Lucas, Sankar Das Sarma, and Alex Levchenko for helpful discussions. The major experimental work is supported by ONR MURI (N00014-21-1-2377). P.K. acknowledges support from DOE (DE- SC0012260). B.S. was supported by the NSF under Grant No. DMR-2045742. K.W. and T.T. acknowledge support from the JSPS KAKENHI (Grant Numbers 21H05233 and 23H02052) and World Premier International Research Center Initiative (WPI), MEXT, Japan. Nanofabrication was performed at the Center for Nanoscale Systems at Harvard, supported in part by an NSF NNIN award ECS- 00335765.

# Supplementary Materials

Materials and Methods

Supplementary Text

Figs. S1 to S4

Supplementary References

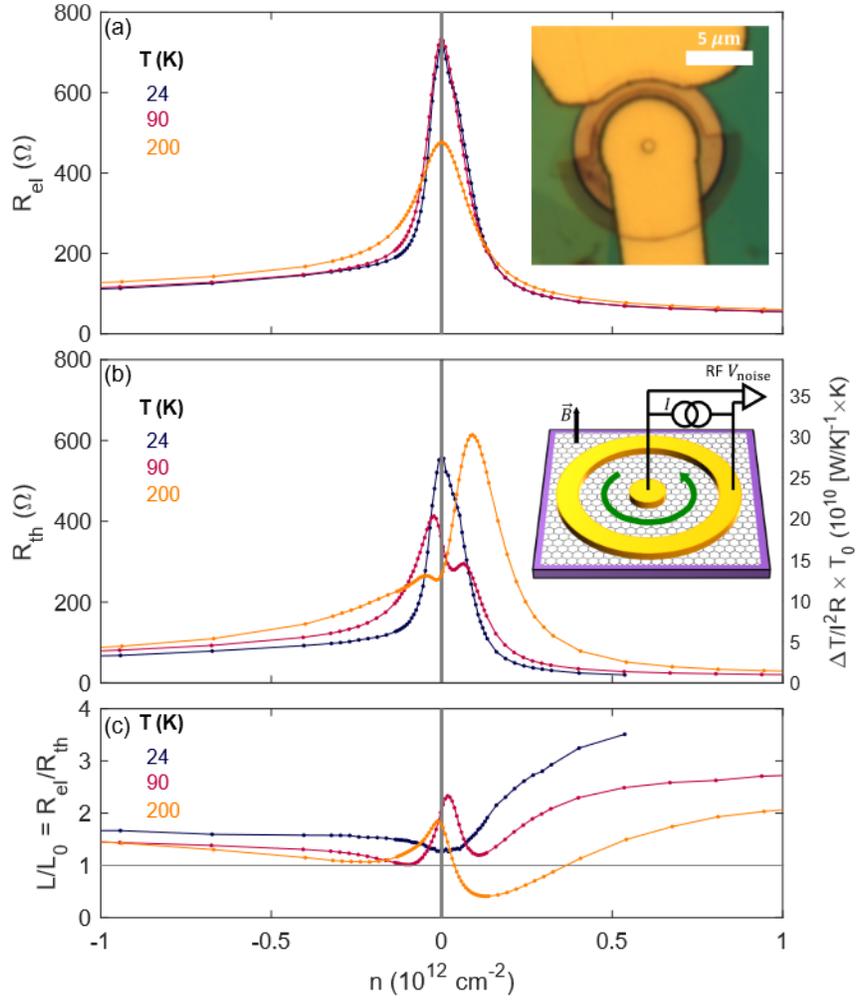

Figure 1: | **Characterization of Corbino devices at** $B = 0$. **(a)** 2-terminal electrical resistance $R_{el}$ as a function of charge density $n$ at three representative temperatures. Inset: optical microscope image of a typical Corbino device. In this device the center contact overhangs the entire channel to provide a homogeneous contact-induced gating effect. **(b)** Thermal resistance $R_{th}$, scaled by the bath temperature (right axis), and converted to $\Omega$ units via the WF law for comparison to electrical resistance (left axis). At higher temperature, the shape qualitatively deviates from that of the electrical resistance. Inset: Schematic diagram of Corbino device and measurement circuit showing DC bias current and RF noise measurement. Magnetic field is applied perpendicular to the graphene plane inducing vortical currents (green arrow). **(c)** Lorenz ratio obtained from the ratio of electrical to thermal resistance divided by the Lorenz constant. At higher temperatures, a peak appears at $n = 0$ and a suppression appears away from the charge neutrality.

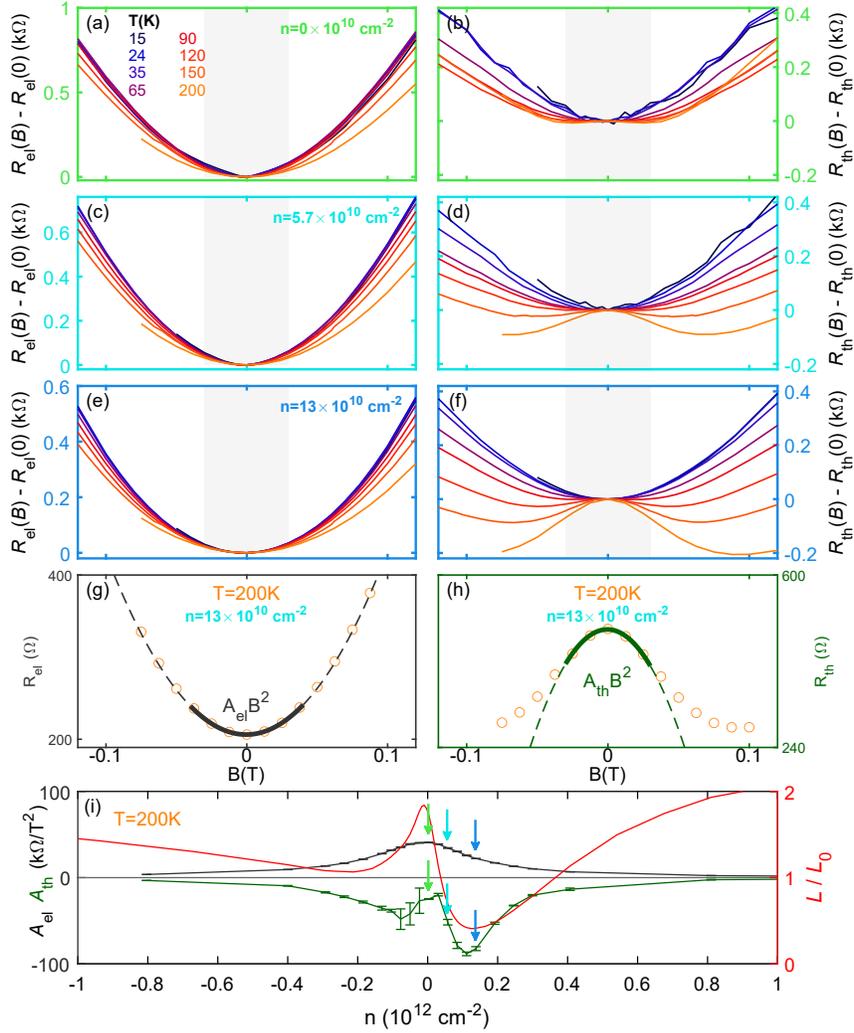

Figure 2: | **Electrical and thermal magneto-resistance and parabolic fits.** (**a,b**) EMR and TMR for $T_{bath} = 15 - 200$ K at $n = 0$. Gray shaded boxes indicate $|B| < 30$ mT fitting range for quadratic functions. (**c,d**) For $n = 5.7 \times 10^{10}$ cm$^{-2}$. (**e,f**) For $n = 13 \times 10^{10}$ cm$^{-2}$. The EMR is positive and parabolic at all temperatures and densities while the TMR is negative at low $B$ for certain temperatures and densities. (**g**) Representative EMR parabolic fit for 200 K to extract the $B^2$ prefactor $A_{\mathrm{el}}$. Circles: data, solid line: parabolic fit, dashed line: extrapolated quadratic functions. The data continues to follow a parabolic trend up to the measured $B = 200$ mT. (**h**) TMR fit to extract the $B^2$ prefactor $A_{\mathrm{th}}$, analogous to (g). The data deviates from parabolic behavior beyond $B \sim 30$ mT. (**i**) Left axis: $A_{\mathrm{el}}$ and $A_{\mathrm{th}}$ at $T_{bath} = 200\,K$. Error bars indicate statistical $1\sigma$ uncertainty. Colored arrows (light green, cyan, light blue) match the corresponding densities plotted in (a-f). Right axis: overlaid $L/L_0$ at $B = 0$ (right axis). $L/L_0 < 1$ occurs at approximately the same density as the most negative $A_{\mathrm{th}}$ coefficient.

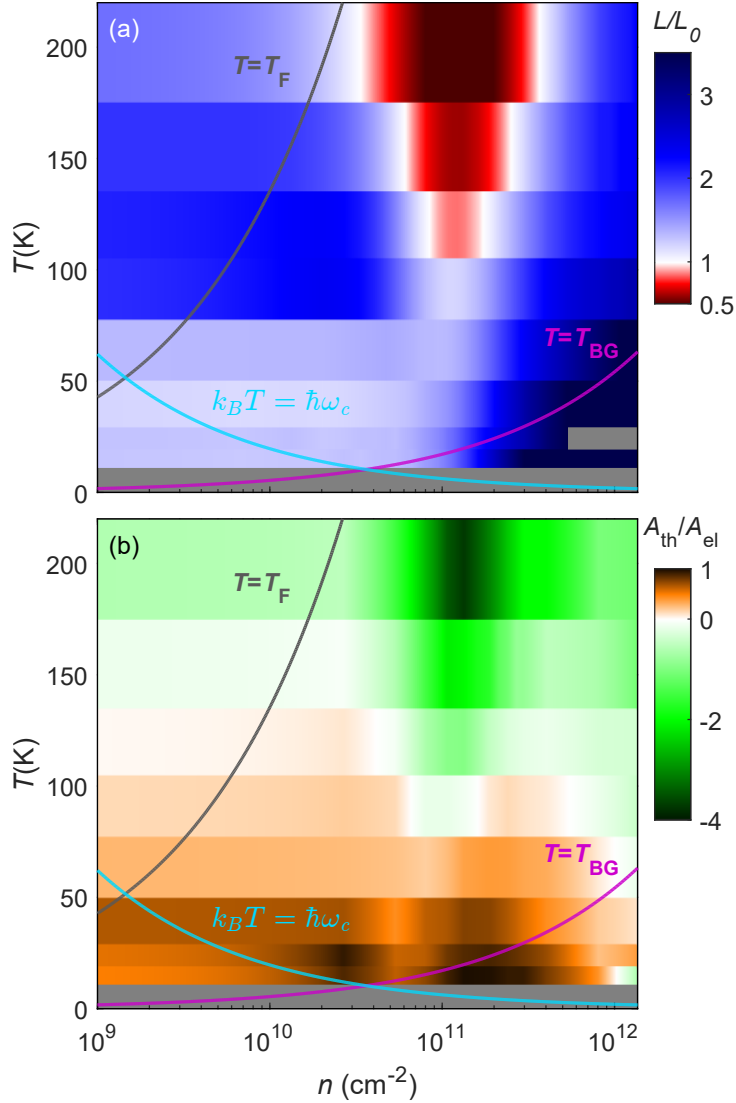

Figure 3: | **Correlation of zero-field Lorenz ratio suppression and negative thermal magneto-resistance coefficient.** (a) Colormap of $L/L_0$ at $B = 0$ versus $n$ and $T$. White corresponds to $\mathcal{L}/\mathcal{L}_0 = 1$. Positive densities (electron filling) are shown. For $T \lesssim 90K$ and $n \lesssim 10^{11} cm^{-2}$, $L/L_0 \sim 1$ and gate-independent, consistent with the WF regime. At higher densities, the Lorenz enhancement appears due to ballistic transport dominated by contact resistance, where the thermal measurement of the channel fails (See SM). For $T \gtrsim 90K$, $L/L_0 < 1$ around $n = 1.3 \times 10^{11}$ cm$^{-2}$. Gray line: Fermi temperature $T = T_F$. Green line: Bloch-Gruneisen temperature $T = T_{BG}$. Red line: temperature scale of Landau level spacing $k_B T = \hbar \omega_c$. (b) Colormap of the TMR to EMR coefficient ratio, $A_{th}/A_{el}$. White corresponds to $A_{th}/A_{el} = 1$. The region of negative $A_{th}/A_{el}$ corresponds with the region of $L/L_0 < 1$ in (a). Lines as described in (a).

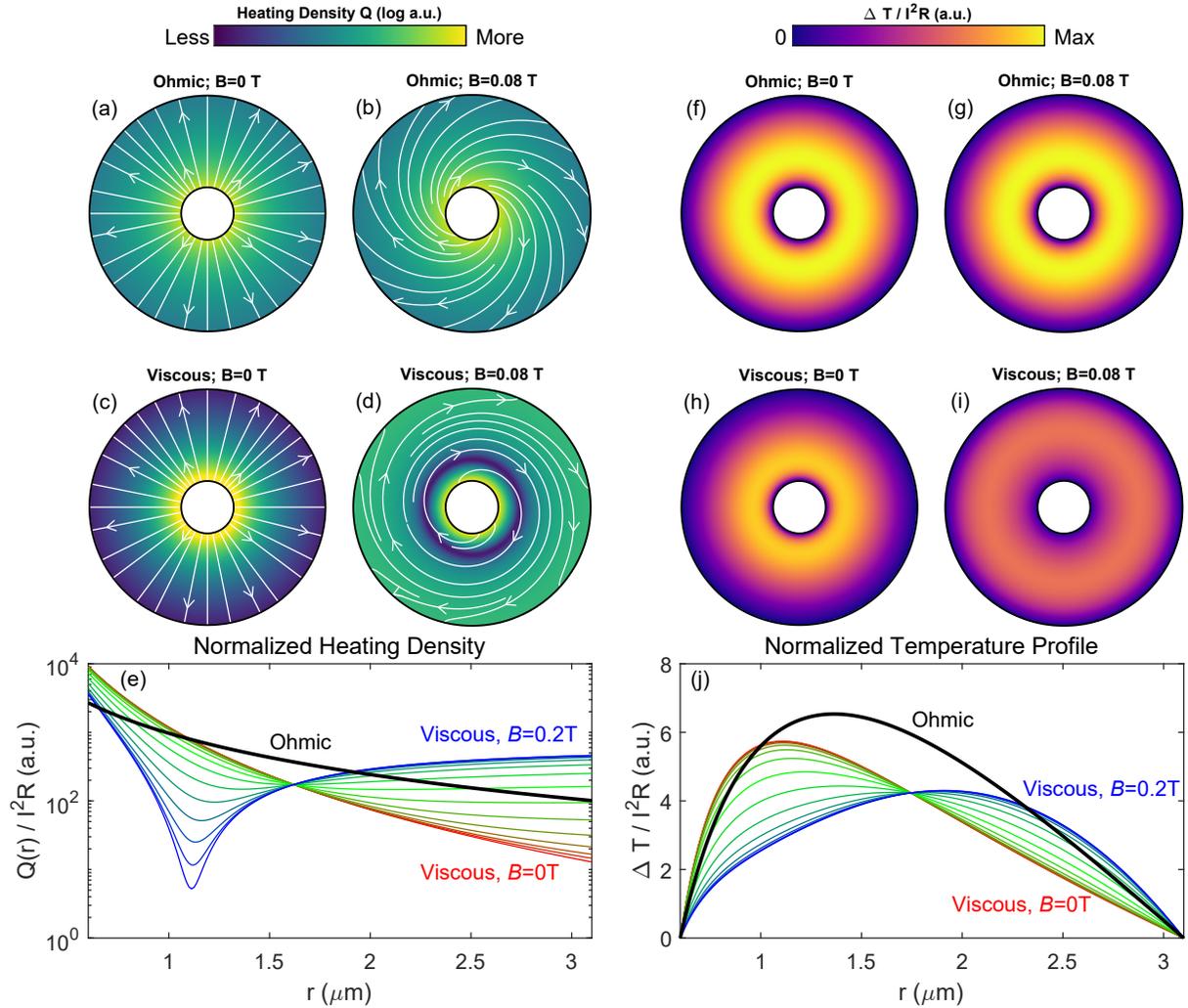

Figure 4: | **Calculated heating and temperature profiles for Ohmic and viscous Corbino transport.** **(a-c)** Calculated heating density $Q$ for (a) Ohmic at $B = 0$ (b) Ohmic at $B = 80$ mT (c) viscous at $B = 0$ (d) viscous at $B = 80$ mT. White streamlines indicate flow velocity. **(e)** Linecuts of normalized heating density $Q/I^2R$. Black line: Ohmic (independent of $B$). Blue through red lines: viscous for $B = 0 \to 200$ mT. **(f-i)** Calculated temperature change distribution $\Delta T$ for (f) Ohmic at $B = 0$ (g) Ohmic at $B = 80$ mT (h) viscous at $B = 0$ (i) viscous at $B = 80$ mT. **(j)** Linecuts of normalized temperature change profile $\Delta T/I^2R$. Black line: Ohmic (independent of $B$). Blue through red lines: viscous for $B = 0 \to 200$ mT.

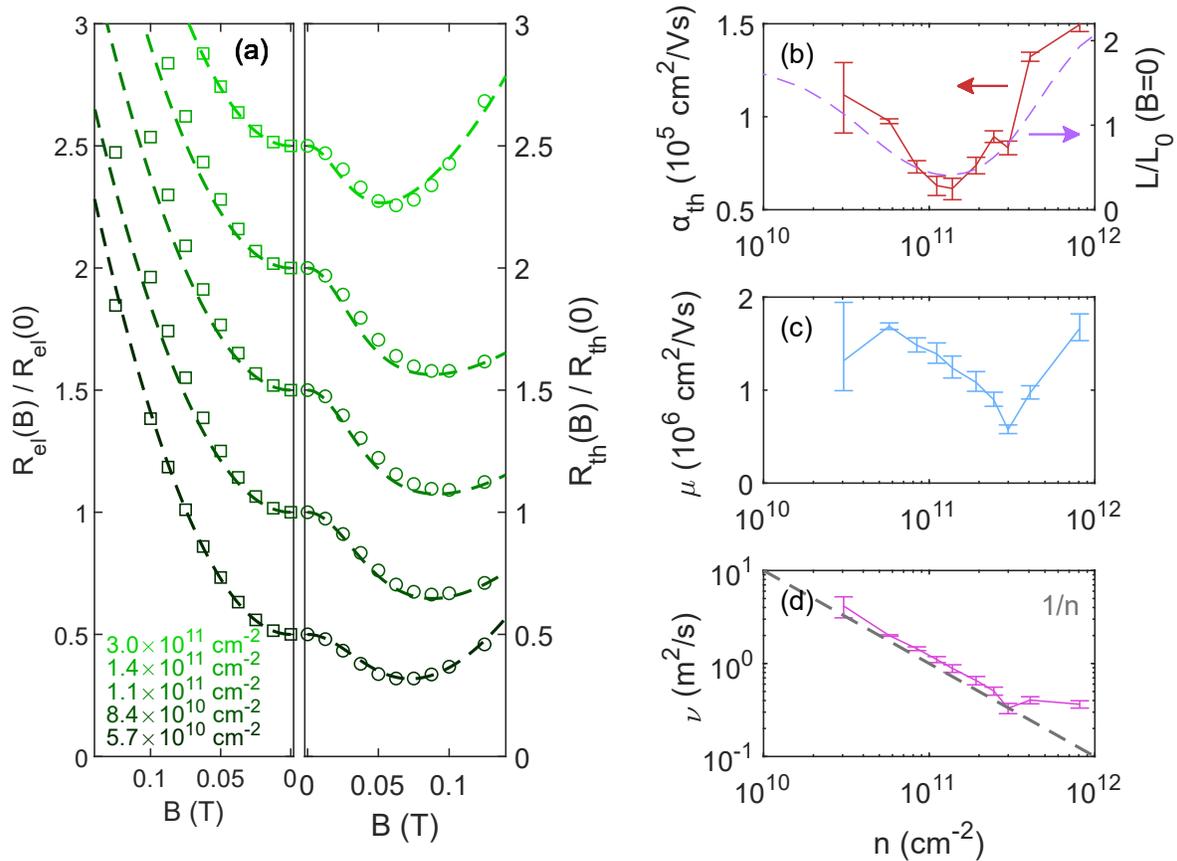

Figure 5: | **Theory Fitting for Thermal and Electrical Mobilities and Viscosity.** (a) Left panel: Measured electrical magnetoresistance (EMR) (square symbols), at $T_{bath} = 200K$, for densities indicated by the color-coded insert text (lower left). Right panel: Measured thermal magnetoresistance (TMR) (circle symbols), at densities and temperature identical to the left panel. Both panels: Dashed lines are theory fits (see main text), and curves are offset by $0.5$ for visibility. (b) Red solid line (left axis, indicated by red arrow): Thermal mobility $\alpha_{th}$ obtained from the fits in panel (a) (see main text). Dashed purple line (right axis, indicated by purple arrow): Lorenz ratio at zero magnetic field corresponding to data in (a). (c) Electrical mobility $\mu$ obtained from the fits in panel (a). (d) Viscosity $\nu$ obtained from the fits in panel (a). Dashed grey line: $1/n$ guide-to-the-eye.

# Supplementary Materials
# for
# Observation of Electronic Viscous Dissipation in Graphene Magneto-thermal Transport

Artem Talanov[1,2]*, Jonah Waissman[1,3]*, Aaron Hui[4], Brian Skinner[4],
Kenji Watanabe[5], Takashi Taniguchi[6], Philip Kim[1,2]*

[1]Department of Physics, Harvard University, Cambridge, 02138, MA, USA
[2]John A. Paulson School of Engineering and Applied Sciences,
Harvard University, Cambridge, 02138, MA, USA
[3]Institute of Applied Physics, The Hebrew University of Jerusalem,
Jerusalem, Israel 9190401
[4]Department of Physics, The Ohio State University, Columbus, 43202, OH, USA
[5]Research Center for Electronic and Optical Materials,
National Institute for Materials Science, 1-1 Namiki, Tsukuba, 305-0044, Japan
[6]Research Center for Materials Nanoarchitectonics,
National Institute for Materials Science, 1-1 Namiki, Tsukuba, 305-0044, Japan

*Corresponding author. Email: pkim@physics.harvard.edu

---

*These authors contributed equally to this work

# Contents

# 1 Materials and Methods

## 1.1 Measuring Thermal Conductance in Self-Heating

We measure thermal conductance using Johnson noise thermometry in a self-heating configuration (*1, 2*), which we summarize here. Graphene electrons are Joule heated by passing a low-frequency (typically $\sim 13$ Hz) AC current through a two-terminal graphene device. The bias heating current is applied with a Stanford Research Systems DS360 Ultra-low distortion function generator operating in balanced mode and outputting a sine voltage excitation, with two series precision external bias resistors on each side of the sample, typically $100 k\Omega$ each. The sample resistance is much lower than the bias resistors, creating an effective ultra-low distortion sinusoidal current bias.

To measure $\Delta T_{avg}$, we use high-bandwidth Johnson noise thermometry (*1, 2*). For the magnetotransport measurements shown here, the noise measurement is implemented in a variable-temperature insert (VTI) cryostat with a superconducting magnet, and the cryogenic low-noise amplifiers (LNAs) sit in a 77K liquid nitrogen dewar external to the VTI cryostat to avoid magnet interference and temperature variations.

The high-impedance graphene channel is differentially impedance-matched to 100 Ohms and the emitted noise is amplified with two wideband RF amplification stages over ∼100-500 MHz bandwidth (*?, 2*). We use a bath-temperature-sweep to calibrate the gain of the amplification chain for all measured resistances, allowing us to convert between Johnson noise and electron temperature for any sample resistance. The LNA provides about 30dB of gain and a secondary room temperature amplifier provides an additional 50dB of gain. The LNAs are used as a differential pair and are tuned to have the same gain in the noise measurement bandwidth. A Herotek zero-bias Schottky diode power detector converts the incoming amplified RF noise power to a quasi-DC voltage, which we measure with a Keysight DMM and a Stanford SR830 lock-in amplifier. The temperature of the sample is modulated at twice the bias current frequency, and the lock-in amplifier measures the $2f$ modulation of the quasi-DC power diode output voltage.

The generalized thermal conductance in units of W/K is computed as the ratio of the the total applied Joule power $I^2 R$ to the temperature rise $\Delta T_{avg}$. As long as the thermal conductivity $\hat{\kappa}$ is proportional to the electrical conductivity $\hat{\sigma}$ and Ohm's law holds, this ratio was shown to be related to the thermal conductance via a universal numerical prefactor, 12, for any two-terminal geometry (*3*).

## 1.2 Thermoelectric Transport in the Corbino geometry

The thermoelectric transport matrix relating electric fields, temperature gradients, and charge and heat current densities, can be used to determine the relationship between an imposed energy current and resulting longitudinal temperature gradient. In a rectangular geometry in non-zero magnetic field, the open transverse boundary conditions leads to a thermal response that includes on- and off-diagonal components of the thermoelectric coefficients.

In the Corbino geometry, the vortical component $E_\theta$ of the electric field must be zero; thus, measuring the two-terminal electrical resistance of the device, even at non-zero $B$, gives exclusively the on-diagonal elements of the conductivity tensor, $\sigma_{rr}$, with $J_r = \sigma_{rr} E_r$, where $J_r, E_r$ are the radial components of the current density and electric field. Likewise, the vortical component $\Delta T_\theta$ of the temperature gradient must be zero, removing the transverse, or Nernst, thermoelectric coefficients.

## 1.3 Sample Fabrication

The graphene Corbino devices are fabricated using standard polymer-assisted van der Waals heterostructure dry assembly techniques (*4*) and microfabrication fabrication procedures with e-beam lithography, etching, and metal deposition (*5*). The central contact is connected to a lead insulated from the rest of the structure by an hydrogen silsesquioxane (HSQ) resist bridge.

## 1.4 Lorenz Ratio in the Ballistic Regime at $B = 0$

Here, we explain how dominant contact resistance in the ballistic regime can lead to artificially enhanced Lorenz ratios in self-heating Johnson noise thermometry.

The total sample resistance can be written as $R_s + 2R_c$, where $R_s$ is the main channel resistance and $R_c$ is each contact resistance. For simplicity, we assume that each contact resistance is point-like, hence contributing no temperature gradient. The Joule power dissipated in the device is then $P = I^2 (R_s + 2R_c)$, leading to a generalized thermal conductance of

$$G_{th,gen} = P/\Delta T = \frac{I^2 (R_s + 2R_c)}{\Delta T}, \tag{1}$$

The contact resistance thus serves to enhance the measured thermal conductance, and in turn the Lorenz ratio. However, its quantitative effect depends on the measured $\Delta T$, which depends on the proportion of applied power dissipated in the channel, as opposed to the contact region. Furthermore, whether the contact resistance generates a temperature gradient or leads to full equilibration of the temperature distribution with the contact is unknown, and can also change $\Delta T$. Simple models show that in either case, the measured Lorenz number is enhanced (see Supplementary Information). This is borne out by self-heating measurements in the quantum Hall regime, in which plateau regions dominated by ballistic edge currents exhibit orders of magnitude enhancement of the measured Lorenz ratio (*6*).

## 1.5  Calculation of Viscosity and Momentum Relaxation Rate

The reshaping of the temperature profile due to viscosity can act to decrease $R_{th}$ for a given fixed $\kappa_{xx}$. Our theoretical analysis shows that the net effect of viscous dissipation and hydrodynamic noise generation modifies the generalized thermal resistance, which can be written as

$$R_{th} = \frac{1}{12\kappa_{xx}} \frac{\ln(r_o/r_i)}{2\pi} \times f(B, \nu, \gamma_{MR}). \qquad (2)$$

where $\nu$ is the kinematic viscosity and $\gamma_{MR}$ is the momentum-relaxing scattering rate (*7*). The scaling function $f(B, \nu, \gamma_{MR})$ possesses two analytic limits. The first is fully Ohmic, in which $\nu \to 0$. The second is fully viscous, and is defined by $\lambda \to \infty$, where the viscous parameter $\lambda = \frac{l_G}{r_i}$, with the radius of the inner contact $r_i$ and the Gurzhi length $l_G = \sqrt{\frac{\nu}{\gamma_{MR}}}$. In the mixed Ohmic-viscous regime it may be computed numerically (*7*). Importantly, the viscous limit is distinct from the hydrodynamic limit $\frac{\gamma_{ee}}{\gamma_{MR}} \gg 1$ that leads to Lorenz suppression at $B = 0$. This is because the viscosity is a decreasing function of $\gamma_{ee}$, so that for very large $\gamma_{ee}$, the Stokes-Ohm equation recovers the purely Ohmic form and viscous effects vanish ($\lambda \to 0$) despite the dominant electron-electron scattering.

Using the scaling function $f$, the thermal magnetoresistance can be re-expressed as

$$\frac{R_{th}(B)}{R_{th}(0)} = \left[\frac{\kappa_{xx}(B)}{\kappa_{xx}(0)}\right]^{-1} \times \frac{f(B)}{f(0)}. \qquad (3)$$

This form shows that two terms can contribute to $\Delta R_{th}(B)$: the ratio of thermal conductivities, and the ratios of scaling functions. The first term is expected to be an increasing function of $B$ since thermal conductivity, like electrical conductivity, decreases with magnetic field, including in the hydrodynamic regime (*8, 9*). The second term, however, due to viscous reshaping of the heating and temperature profiles in the Corbino geometry, can lead to a decrease in the generalized two-terminal thermal resistance. This happens when the viscous parameter $\lambda$ is sufficiently large.

We calculate the low-$B$ electrical magnetoresistance coefficient $A_{el}$ from

$$R_{el}(B) = R_{el}(0) + A_{el}B^2. \tag{4}$$

Ref. 10 provides the formula for Corbino magnetoresistance in the full viscous-Ohmic crossover:

$$\Delta R_{el}(B) = \frac{B^2 \ln a}{2\pi \rho_0 (ne)^2} \times F(a,b), \tag{5}$$

where the function $F(a,b)$ depends on two parameters, $a = r_o/r_i$ and $b = r_o/l_G$.

The EMR and TMR provide a system of two equations from which we obtain $(\nu, \gamma)$. To calculate the error bars, we numerically solve for $(\nu, \gamma)$ again using all 9 combinations of $(A_{el} \pm \Delta A_{el}, \alpha_{th} \pm \Delta \alpha_{th})$, where $\Delta A_{el}$, $\Delta \alpha_{th}$ are the 68% confidence intervals from the quadratic fits of the $R_{el}(B)$, $R_{th}(B)$ data. The lowest and highest $(\nu, \gamma)$ from the 9 results are taken as the confidence interval in Fig. **??**(e).

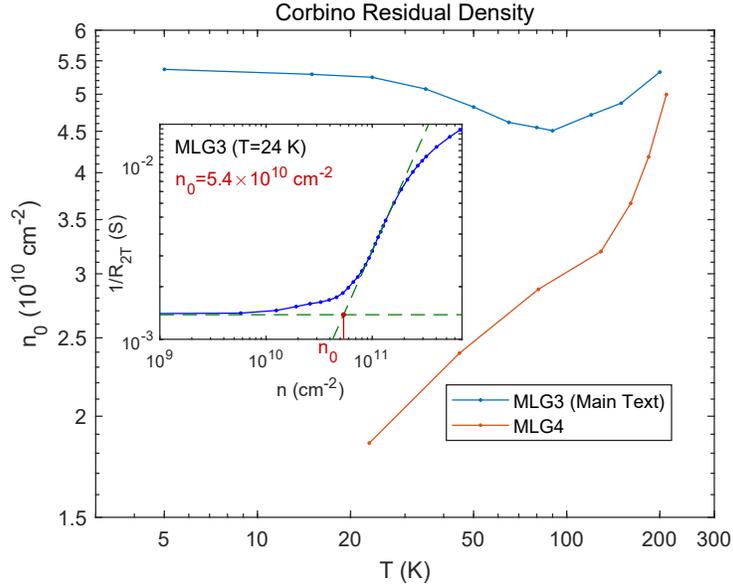

Figure S1: Main panel: Residual density of both devices in this work. Inset: log-log plot of electrical conductance vs $n$ (blue). Extrapolations to the linear and flat parts are indicated by the green dashed lines. Red indicates the intersection of these, which is the residual density $n_0$.

## 2 Supplementary Text

### 2.1 Device Disorder and Residual Density

Here, we present measurement of the device disorder level. Figure S1 shows the residual density versus temperature for both of the devices exhibited in this work.

## 2.2 Results for an Additional Device

In this section we present data from a second monolayer graphene Corbino device. For this device, the measured Lorenz ratio does not show an absolute suppression $L/L_0 < 1$, only a relative suppression. Nonetheless, the relative Lorenz suppression is observed in the degenerate $T < T_F$, quasi-elastic $T > T_{BG}$ region of parameter space, similar to the device of the main text. This device has a shorter channel length of $\sim 1.64$ $\mu$m. The shorter channel length and smaller radii compared to the device shown in the main text result in a larger region of expected viscous behavior, since the device dimensions are even smaller than the Gurzhi length $l_G$ compared to the first device. Figure S2 shows the two devices, and compares their electrical resistance and Lorenz ratio versus density for varying temperature. We observe similar Dirac peak behavior, and comparable Lorenz ratio behavior with the zero density peak, relative suppression for moderate densities, and enhancement at large densities (see main text and below). Figure S3 shows the electrical and thermal magnetoresistance (EMR and TMR) parabolic prefactors $A_{el}$ and $A_{th}$, respectively, versus density for selected temperatures, compared with the zero-field Lorenz ratio. Here, we observe a similar emergence of negative TMR corresponding with the relative suppression of the zero-field Lorenz ratio in density and temperature. Figure S4 shows zero-field Lorenz ratios and the parabolic prefactor ratio $A_{th}/A_{el}$ versus density and temperature, showing the emergence of the relative Lorenz suppression and negative TMR regions in the degenerate and quasi-elastic regime, as seen in the device discussed in the main text.

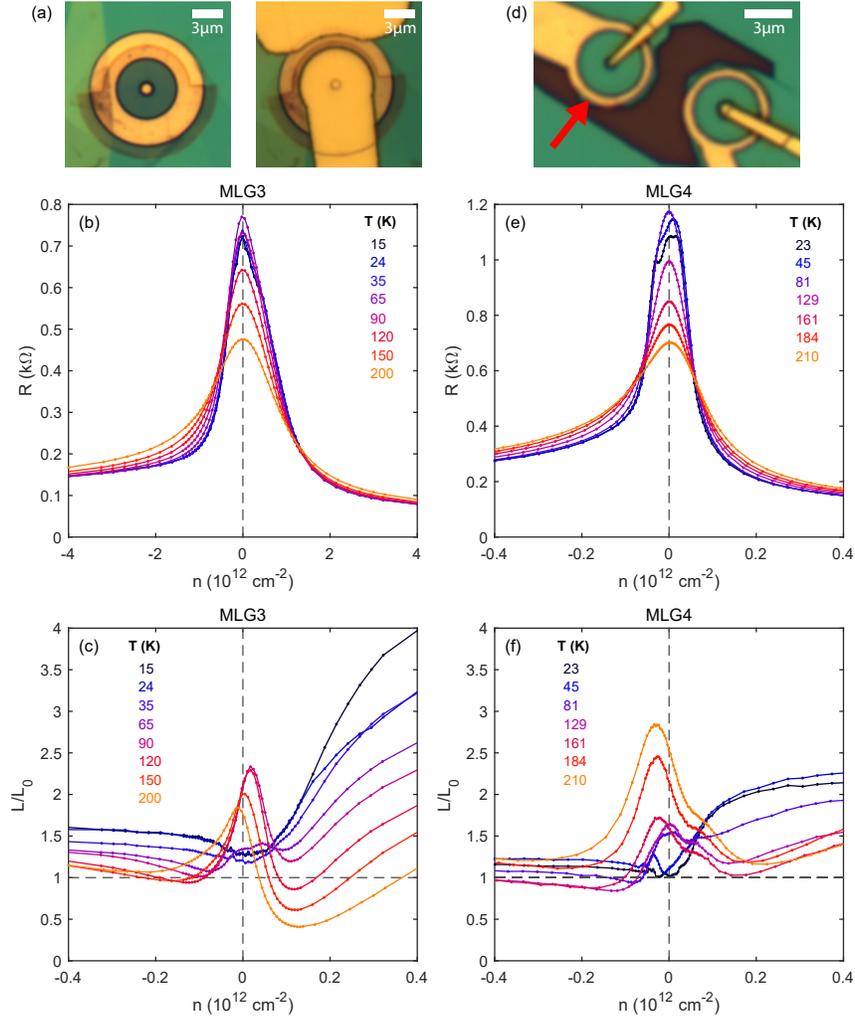

Figure S2: a) Optical micrograph of the device shown in the main text (MLG3), before (left) and after (right) deposition of the central contact lead. b) Electrical resistance $R$ of MLG3 versus density $n$ for varying bath temperatures $T$. c) Lorenz ratio $L/L_0$ versus $n$ for varying $T$. d) Optical micrograph of a second Corbino device (MLG4) (left device, red arrow). e) $R$ versus $n$ for varying $T$ for MLG4. f) $L/L_0$ versus $n$ for varying $T$ for MLG4.

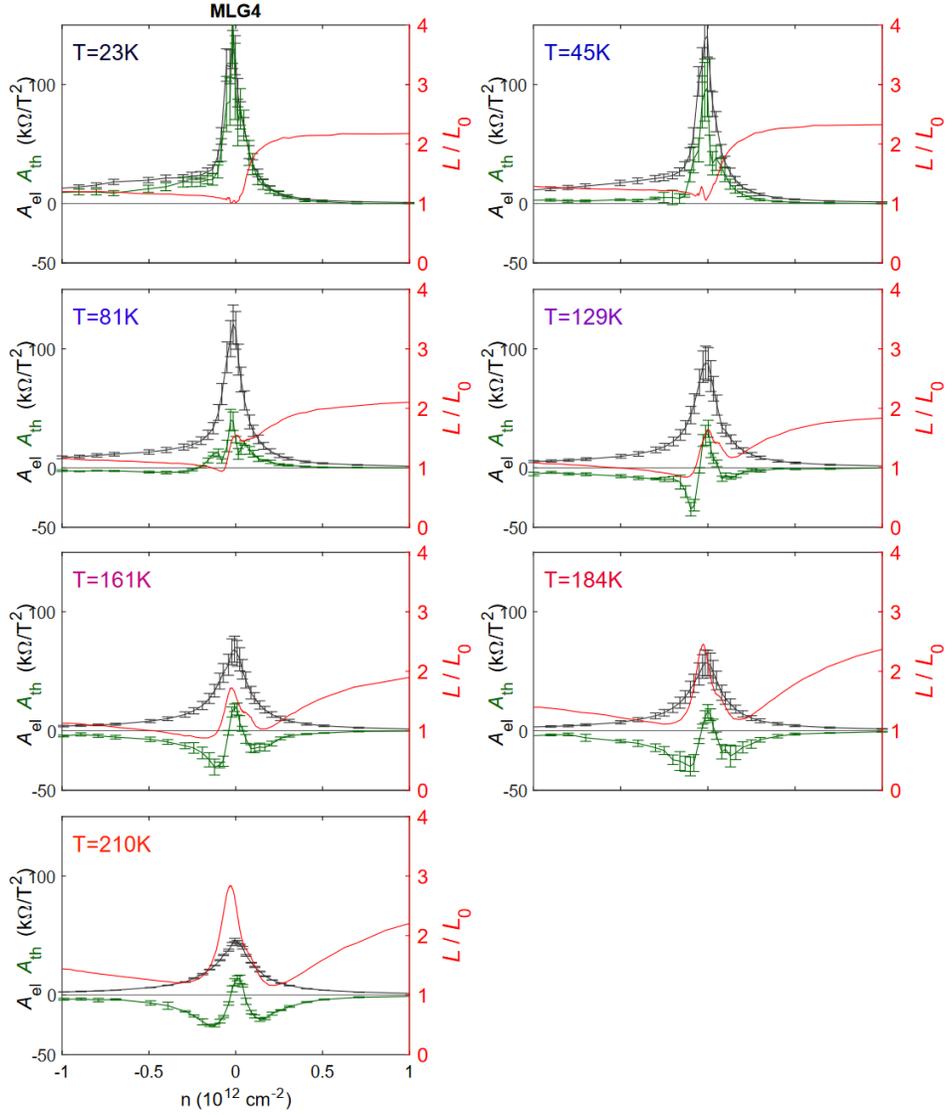

Figure S3: Left axes: Line cuts of $A_{el}$ (dark green) and $A_{th}$ (light green) versus $n$ for varying $T$ from additional device MLG4. Right axis: Line cuts of $L/L_0$ versus $n$. See main text.

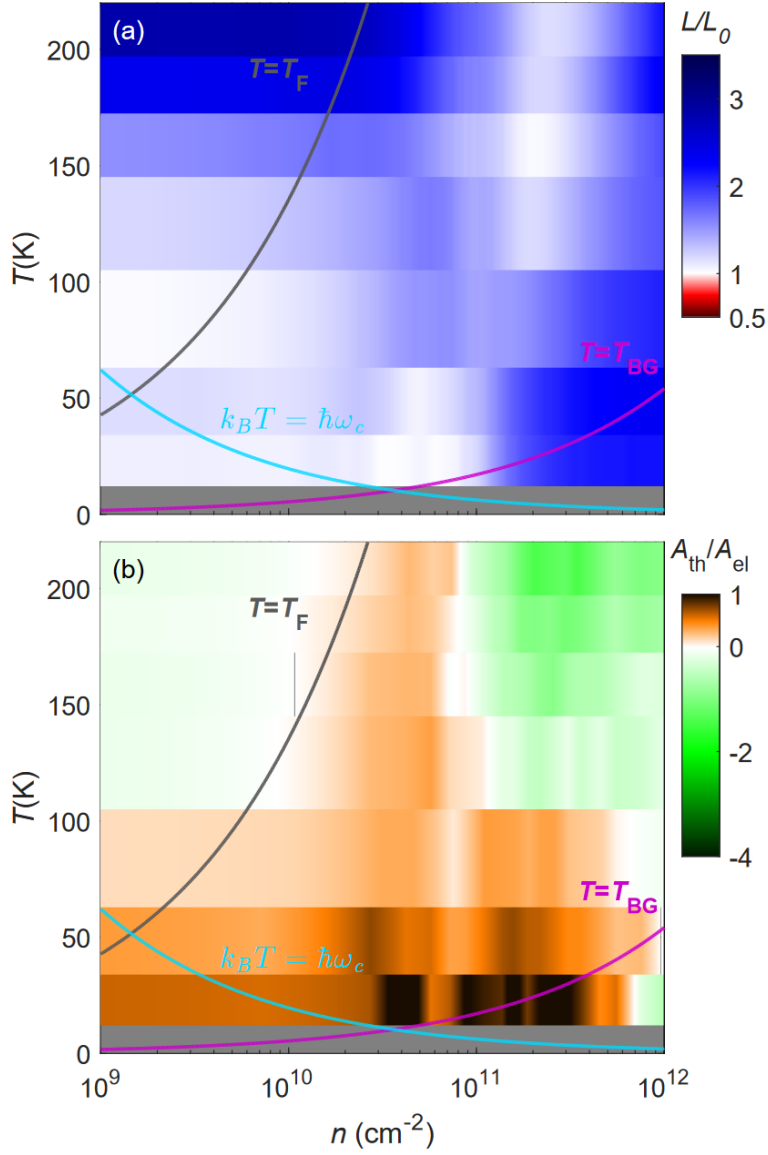

Figure S4: a) Lorenz ratio $L/L_0$ versus density $n$ and temperature $T$ for additional device MLG4 at $B = 0$. Grey line: $T = T_F$ the Fermi temperature. Magenta line: $T_{BG}$, the Bloch-Gruneisen temperature. Cyan line: cyclotron energy in temperature units. See main text. b) Ratio of thermal magnetoresistance curvature $A_{th}$ to electrical magnetoresistance curvature $A_{el}$ versus $n$ and $T$.

## 2.3 Lorenz Ratio in the Ballistic Regime at $B = 0$ in a Contact Resistance Model

In the ballistic regime, our self-heating Johnson noise thermometry measurement technique produces artificially enhanced Lorenz ratios. This is different from the behavior predicted for a diffusive channel whose resistance is dominated by contact resistance, where the measured Lorenz ratio would trend towards the Lorenz ratio of the contact resistance itself, presumed to be $\frac{\mathcal{L}_c}{\mathcal{L}} = 1$ (*11*). Here, we present a more generalized version of the contact resistance model developed in Ref. 11.

The total sample resistance is $R_s + 2R_c$, where $R_s$ is the main channel resistance and $R_c$ is each contact resistance. For simplicity, we assume that each contact resistance is point-like, and thus there is no temperature gradient in the contact. We use $T_0$ to denote the bath temperature.

We wish to obtain a formula that gives us the measured Lorenz number $\mathcal{L}_m$ as a function of the sample Lorenz number $\mathcal{L}_s$ and the contact resistance Lorenz number $\mathcal{L}_c$, as well as other fixed or known parameters in the system including resistance. Most generally, we can write the generalized thermal conductance of the sample, using the total Joule power $P$ and the average Johnson noise temperature rise $\overline{\Delta T}$, as

$$G_{th,gen} = P/\overline{\Delta T} = \frac{I^2 \left(R_s + 2R_c\right)^2}{R_s \overline{\Delta T_s} + 2R_c \overline{\Delta T_c}}, \tag{6}$$

where $\overline{\Delta T_s}$ and $\overline{\Delta T_c}$ are respective average Johnson noise temperature rises of the sample channel and contact above $T_0$.

We define the contact resistance $R_c$ to have a thermal resistance $R_{c,th} = 1/G_{th,c}$ such that the temperature drop $\Delta T_{cs}$ across the contact resistance is related to the heat $Q_c$ that flows through it via

$$\Delta T_{cs} = \frac{Q_c}{G_{th,c}}, \tag{7}$$

and we can likewise define a contact Lorenz number as

$$\mathcal{L}_c = \frac{G_{th,c} R_c}{T_0}. \tag{8}$$

We will assume, without rigorous justification, that the average Johnson noise temperature of the contact resistance is one half of $\Delta T_{cs}$. This would be the case for a spatially extended uniformly resistive contact resistance, and as in Ref. 11, we use the same model for a point-like contact resistance.

Ref. 11 assumed that the Joule heat generated at each contact is distributed equally on each side of the contact resistance; the half that goes into the metal contact is immediately thermalized to bath temperture, and the half that goes into the graphene channel flows through the contact resistor and causes a temperature rise. Here, instead of assuming that the contact resistance Joule power is equally distributed on each side, we will instead suppose that a fraction $\alpha$ is distributed to the graphene side, and $1 - \alpha$ is distributed towards the metal contact. In the case of a ballistic device, we typically expect $\alpha \to 0$ becaue the dissipation happens inside the metal contact (reservoir) itself after the electrons leave the sample.

We will write $\overline{\Delta T_c} = \frac{1}{2} \Delta T_{cs}$ in terms of the total electrical current and the contact Lorenz number. The heat current $Q_c$ flowing through each contact resistance is then half of the heat generated in the channel $R_s$ added to the fraction $\alpha$ of heat generated in the contact resistance $R_c$:

$$Q_c = \frac{1}{2} I^2 R_s + \alpha I^2 R_c. \tag{9}$$

We then obtain

$$\overline{\Delta T_c} = \frac{1}{2} \Delta T_{cs} = \frac{1}{2} \frac{Q_c}{G_{th,c}} = \frac{1}{2} \frac{\frac{1}{2} I^2 R_s + \alpha I^2 R_c}{\mathcal{L}_c T_0 / R_c} \Delta T_{cs} = \frac{1}{2} I^2 \frac{R_c}{\mathcal{L}_c T_0} \left( R_s + 2\alpha R_c \right). \tag{10}$$

The effective bath temperature for the channel becomes $T_0 + \Delta T_{cs}$. We can define an effective self-heating average Johnson noise temperature rise for just the channel, above the effective

bath temperature, as

$$\widetilde{\Delta T_s} = \overline{\Delta T_s} - \Delta T_{cs} = \frac{I^2 R_s}{G_{th,gen,s}} = \frac{I^2 R_s}{12 T_0 \mathcal{L}_s / R_s} \tag{11}$$

where $G_{th,gen,s}$ is the generalized thermal conductance of the sample channel only, using the sample Lorenz ratio $\mathcal{L}_s = \frac{G_{th,gen,s} R_s}{12 T_0}$. This allows us to write

$$\overline{\Delta T_s} = \Delta T_{cs} + \frac{I^2 R_s^2}{12 T_0 \mathcal{L}_s}. \tag{12}$$

Substituting Eqs. 10 into 12 and into 10, and the results into Eq. 6, and re-arranging, we obtain

$$G_{th,gen} = \frac{(R_s + 2R_c)^2}{\frac{R_s^3}{12 T_0} \frac{1}{\mathcal{L}_s} + \frac{R_c}{2 T_0}(R_s + 2\alpha R_c)(R_s + R_c)\frac{1}{\mathcal{L}_c}}. \tag{13}$$

Finally, relating $G_{th,gen}$ to the measured Lorenz ratio as

$$G_{th,gen} = \frac{12 T_0 \mathcal{L}_m}{R_s + 2R_c}, \tag{14}$$

we can write Eq. 13 as

$$\mathcal{L}_m = \frac{(R_s + 2R_c)^3}{\frac{R_s^3}{\mathcal{L}_s} + \frac{6R_c(R_s + 2\alpha R_c)(R_s + R_c)}{\mathcal{L}_c}}. \tag{15}$$

In the ballistic limit, we have $\alpha \to 0$. Microscopically, this corresponds to electrons scattering and randomizing their momenta only once in the large reservoirs (corresponding to the metal contacts in our device) of the ballistic wire model. Applying this limit and re-arranging, we find the measured Lorenz ratio becomes

$$\mathcal{L}_m = \mathcal{L}_s \frac{(R_s + 2R_c)^3}{R_s^3 + 6 R_c R_s (R_s + R_c) \frac{\mathcal{L}_s}{\mathcal{L}_c}}. \tag{16}$$

Expanding Eq. 16 in $\frac{R_s}{R_c} \to 0$ (as occurs in a quasi-ballistic or ballistic system), we obtain to the lowest two orders

$$\mathcal{L}_m \approx \mathcal{L}_c \cdot \frac{4}{3} \frac{R_c}{R_s}\left(1 + \frac{1}{2}\frac{R_s}{R_c}\right), \tag{17}$$

showing how the measured Lorenz ratio will become large in ballistic samples where $R_c \gg R_s$.

# Supplementary References

1. J. Crossno, X. Liu, T. A. Ohki, P. Kim, K. C. Fong, *Applied Physics Letters* **106**, 023121 (2015).

2. A. V. Talanov, J. Waissman, T. Taniguchi, K. Watanabe, P. Kim, *Review of Scientific Instruments* **92**, 014904 (2021).

3. C. Pozderac, B. Skinner, *Physical Review B* **104** (2021).

4. F. Pizzocchero, *et al.*, *Nature Communications* **7** (2016).

5. L. Wang, *et al.*, *Science* **342**, 614 (2013).

6. J. Crossno, Electronic thermal conductance of graphene via electrical noise, Ph.D. thesis, Harvard University (2017).

7. A. Hui, B. Skinner, *Phys. Rev. Lett.* **130**, 256301 (2023).

8. A. Lucas, S. D. Sarma, *Physical Review B* **97** (2018).

9. A. Levchenko, S. Li, A. V. Andreev, *Physical Review B* **106** (2022).

10. A. Levchenko, J. Schmalian, *Annals of Physics* **419**, 168218 (2020).

11. K. C. Fong, Impact of contact resistance in lorenz number measurements (2017).